\documentclass[twoside,british,3p,preprint,number]{elsarticle}
\usepackage[T1]{fontenc}
\usepackage[latin9]{inputenc}
\usepackage{textcomp}
\usepackage{graphicx}
\usepackage{setspace}
\makeatletter
\journal{Cryogenics}

\@ifundefined{showcaptionsetup}{}{%
 \PassOptionsToPackage{caption=false}{subfig}}
\usepackage{subfig}
\makeatother

\usepackage{babel}
\begin{document}

\begin{frontmatter}{}

\title{A lightweight thermal heat switch for redundant cryocooling on satellites}

\author[tm]{M. Dietrich\corref{md}}

\ead{dietrich@transmit.de}

\author[tm]{A. Euler}

\author[tm,iap]{G. Thummes}

\cortext[md]{Corresponding author at: TransMIT-Centre of Adaptive Cryotechnology
and Sensors, Heinrich-Buff-Ring 16, 35392 Giessen, Germany. Tel.:
+49 641 99 33462}

\address[tm]{TransMIT-Centre of Adaptive Cryotechnology and Sensors, 35392 Giessen,
Germany}

\address[iap]{Institute of Applied Physics, University of Giessen, 35392 Giessen,
Germany}
\begin{abstract}
A previously designed cryogenic thermal heat switch for space applications
has been optimized for low mass, high structural stability, and reliability.
The heat switch makes use of the large linear thermal expansion coefficient
(CTE) of the thermoplastic UHMW-PE for actuation. A structure model,
which includes the temperature dependent properties of the actuator,
is derived to be able to predict the contact pressure between the
switch parts. This pressure was used in a thermal model in order to
predict the switch performance under different heat loads and operating
temperatures. The two models were used to optimize the mass and stability
of the switch. Its reliability was proven by cyclic actuation of the
switch and by shaker tests.
\end{abstract}

\tnotetext[T1]{©2017. This manuscript version is made available under the CC-BY-NC-ND
4.0 license http://creativecommons.org/licenses/by-nc-nd/4.0/}

\end{frontmatter}{}


\section{Introduction}

Redundancy concepts are of vital importance for long term satellite
missions \cite{Ross2001}. If cryocoolers are involved, several methods
are possible to achieve redundancy \cite{Ross2005}. One method is
to use two identical cryocoolers; one active and one in stand-by operation.
In the simplest case, both cryocoolers are directly connected to the
cooling load. However, this approach has the drawback that the active
cryocooler has to carry the thermal load introduced by the stand-by
cryocooler, which usually requires an increased power input to the
active part or even a larger cooler model. An increased power input
usually lowers the reliability of the system while a larger cryocooler
model increases the total system mass. An alternative is to use thermal
heat switches \cite{Gilmore2002}, which connect each cooler to a
common heat load (e.g. an infrared detector). This method can reduce
the thermal load of the stand-by cryocooler significantly.

The passively operated thermal heat switch was developed in a previous
project where it was able to demonstrate that the design principle
works \cite{Dietrich2014}. The switch makes use of the very high
coefficient of thermal expansion (CTE) of the thermoplastic Ultra-High
Molecular Weight Polyethylene (UHMW-PE). The high CTE allows for a
relative large gap width of 80 \textmu m in the open state, which
facilitates the manufacturing and enhances the reliability. This first
generation heat switch showed an on-state thermal conductance of 1000
mW/K at 100 K and an off-state thermal conductance of 3 mW/K between
80 and 260 K. However, with a mass of 250 g, the switch was still
too heavy to be launched into space. Additionally, it underwent only
limited reliability testing, especially of the poorly understood creep
behavior of the actuator material. Lastly, the first generation heat
switch performance was adversely impacted by shaker tests.

For further improvement of the switch, a structural and a thermal
model were created with the aim to reduce the mass and to increase
the stability simultaneously. A reduction in on-state conductance
to about 500 mW/K at 100 K was tolerable for the intended cooling
of infrared detectors.

\section{Models and changes in switch design}

\begin{figure}
\begin{centering}
\includegraphics{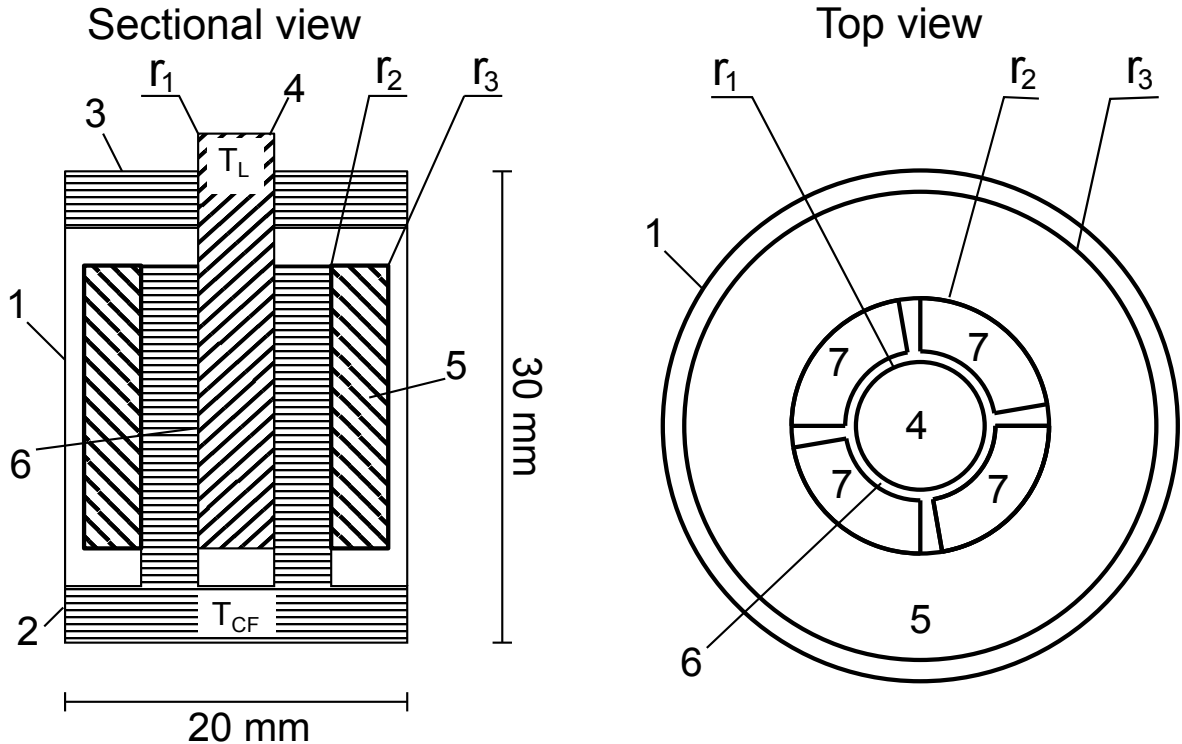}
\par\end{centering}
\caption{Schematic view of the heat switch (not to scale). 1: Connector (titanium
alloy tube), 2: lower switch part with jaws (7) and cold flange ($T_{CF})$
made of aluminium alloy, 3: shaft holder (epoxy or aluminium), 4:
upper switch part with shaft and connection to load ($T_{L})$ made
of aluminum alloy, 5: actuator tube (UHMW-PE), 6: small gap.\label{fig:Schematic-view-of}}
\end{figure}

The principle design of the heat switch, as shown in Fig. 1, is described
in more detail in \cite{Dietrich2014}. The thermal model is used
to estimate the thermal conductance in the on- and off-state. In the
off-state, the main heat flow path is along the mechanical connector
(1) of the two switch parts (2, 4) shown on the left in Fig. \ref{fig:Schematic-view-of}.
 In the old design, this connector consisted of four thin-walled stainless-steel
capillary tubes, which only provided a small structural support. These
capillary tubes were replaced by a single thin-walled tube made of
titanium alloy, which has a lower thermal conduction than the previous
four stainless steel capillaries. To further reduce the thermal conductance
in the off-state, the shaft holder material (3) was changed from copper
to a low outgassing epoxy. Furthermore, the switch width was reduced
in order to lower the total mass. As a consequence of this shrink,
the gap width had to be reduced from 80 \textmu m to 70 \textmu m.
By these measures, the heat conductance in off-state was reduced from
the early 2 mW/K (between 100 and 220 K) to 1.6 mW/K. However, the
thermal model predicted a somewhat higher off-state conductance of
2.3 mW/K. This higher value is probably due to the thermal contact
resistances between the various parts of the real switch, which were
not included in the model.

To be able to calculate the thermal conductance in the on-state, the
thermal model needs information about the pressure between the switch
contact areas. A structural model was therefore developed, which includes
the temperature dependent properties of the materials used, especially
the coefficients of thermal expansion. With regard to the switch geometry,
the contact pressure is then calculated using Hooke's law as
\begin{equation}
P(T)=\frac{E_{al}E_{pe}(r_{2}-r_{3})^{2}(\varepsilon_{al}(T)-\varepsilon_{pe}(T))}{(\nu_{al}-1)\left(E_{al}\nu_{pe}(r_{2}^{2}+r_{3}^{2})-(E_{al}+E_{pe})(r_{2}-r_{3})^{2}\right)},\label{eq:struct_mod}
\end{equation}
where $E,\,\varepsilon,$ and $\nu$ are the creep modulus, thermal
expansion at temperature $T$, and Poisson's ratio, respectively.
The radii $r_{2}$ and $r_{3}$ are indicated in Fig. \ref{fig:Schematic-view-of}
and denote the outer radius of the jaws and the actuator, respectively.
The indices $"al"$ and $"pe"$ stand for the different switch materials
aluminium and UHMW-PE, respectively. Due to the lack of available
data, the CTE of the UHMW-PE actuator material was measured in advance,
which revealed a contraction from 300 K to 100 K of 2.1\% (Al 6061-T6:
0.38\%). The calculated contact pressure at 100 K is a bit lower than
in the first generation design (3.9 MPa vs. 4.2 MPa, primarily due
to the reduction of the switch dimensions. The contact conductance
$h_{c}$ was first estimated from \cite{Bahrami2004} and later measured
using the real switch. The contact conductance $h_{c}$ was calculated
from the experimental results using a model derived from the Fourier-law:
\begin{equation}
\dot{Q}=\frac{\Delta T\,k_{s}h_{c}A}{(r_{1}/2+(r_{2}-r_{1})/2)h_{c}+k_{s}},\label{eq:fourier}
\end{equation}
where $\dot{Q}$ is the heat load, $r_{1}$ and $r_{2}$ are the radii
as indicated in Fig. \ref{fig:Schematic-view-of}, $A$ is the contact
area between jaws and shaft, $\Delta T=T_{L}-T_{CF}$ is the temperature
difference between the connection to the load (4) and cold flange
(2), and $k_{s}$ is the mean thermal conductivity of aluminium of
the jaws and shaft (2, 4).

Measurements of the contact conductance were performed using an in-house
made coaxial pulse-tube cryoccooler (model PT08) employing an SL-400
compressor from AIM GmbH, Heilbronn, Germany \cite{Yang2005}. The
experimental setup with the heat switch mounted n the cold flange
of the PT08 is shown in 
\begin{figure}
\centering{}\includegraphics[height=0.42\textheight]{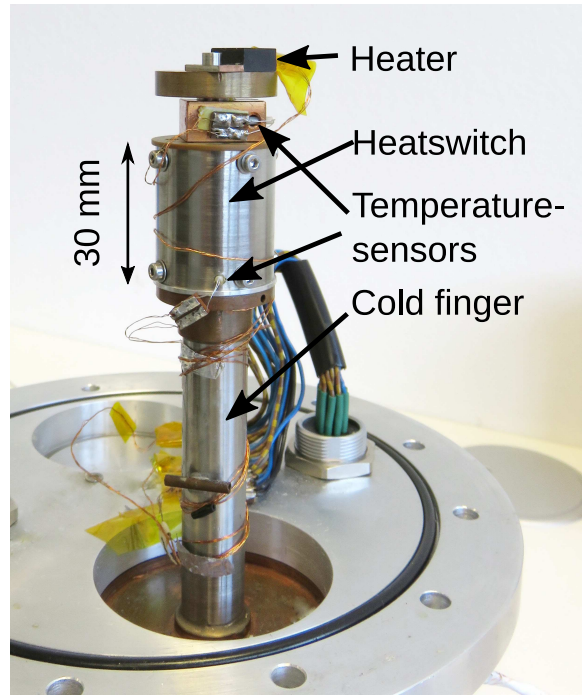}\caption{Photo of the experimental setup.\label{fig:Photo-of-the}}
\end{figure}
Fig. \ref{fig:Photo-of-the}. 

\begin{figure}
\begin{centering}
\includegraphics[width=0.85\textwidth]{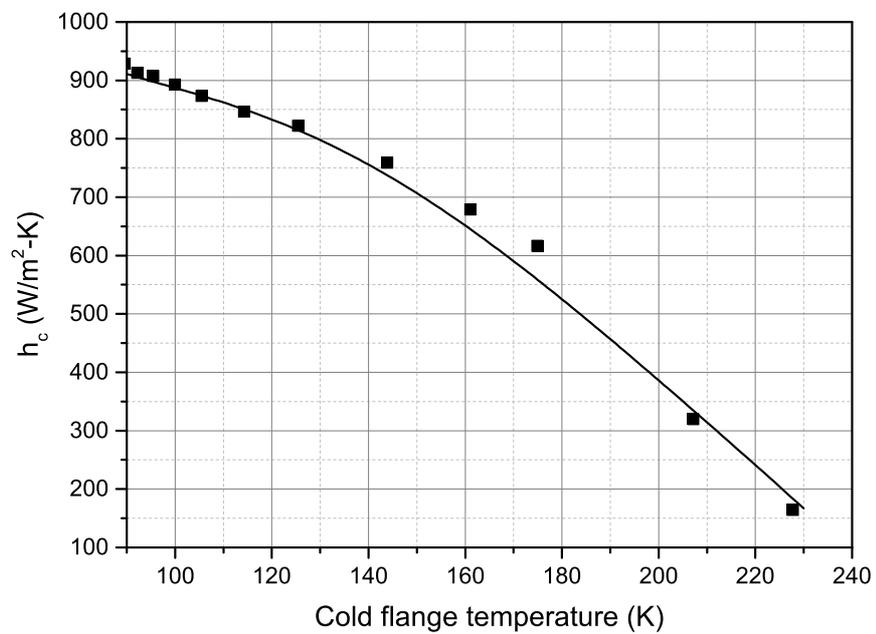}
\par\end{centering}
\caption{Measured (squares) and fitted contact conductance of the heat switch
as function of the cold flange temperature.\label{fig:Measured-and-fitted}}
\end{figure}

The experimental data were then fitted using a power law from \cite{Bahrami2004}:
\begin{equation}
h_{c}=1.25*k_{s}\left(\frac{m}{\sigma}\right)\left(\frac{P}{H}\right)^{0.95}\propto a*P^{0.95}.\label{eq:contact_resistance}
\end{equation}
The single fit parameter $a$ includes the difficult to determine
properties of the switch contact surface, namely the RMS surface roughness
$\sigma$, the RMS slope of the asperities $m$, and the microhardness
$H$, which are assumed to be constant in the relevant temperature
range between 100 and 200 K. The temperature dependent contact pressure
$P$ was calculated from the structural model (see Eq. \ref{eq:struct_mod}).
Fig. \ref{fig:Measured-and-fitted}
 compares the thermal conductance from experimental data with the
contact conductance calculated from Eq. \ref{eq:contact_resistance}.
The fitted model shows a good agreement with the experimental data.
However, the thermal contact conductance $h_{c}$ is a an order of
magnitude lower than predicted by \cite{Bahrami2004}. One reason
is that the microhardness $H$ increases with lower temperature leading
to a smaller contact conductance (see Eq. \ref{eq:contact_resistance}
and \cite{Kumar2004}). Another reason is probably due to the contact
area $A$ that is smaller than assumed. That is, because of the structure
of the switch, both switch halves may not press together everywhere
with the same pressure, especially near the cold flange, where the
jaws may not even touch the shaft. An increase in contact conductance
is therefore anticipated when the contact between shaft and jaws is
better aligned, e.g. by using a slightly conical shaft or flexibly
attached jaws. As to be expected, the measured on-state conductance
of 350 mW/K is lower than the 1000 mW/K reached in the previous project
\cite{Dietrich2014}. Besides the lower contact conductance $h$,
the lower switch on-state conductancecan also be attributed to the
smaller switch dimensions and the use of aluminium instead of copper
as the switch conduction material.

\section{Stability}

In order to verify the structural stability of the new switch design,
it underwent a shaker test, which simulated the vibrations of a launch
vehicle. For the test, two switches were mounted on a test rig simulating
two cold fingers and a sensor carrier 
\begin{figure}
\begin{centering}
\includegraphics[height=0.42\textheight]{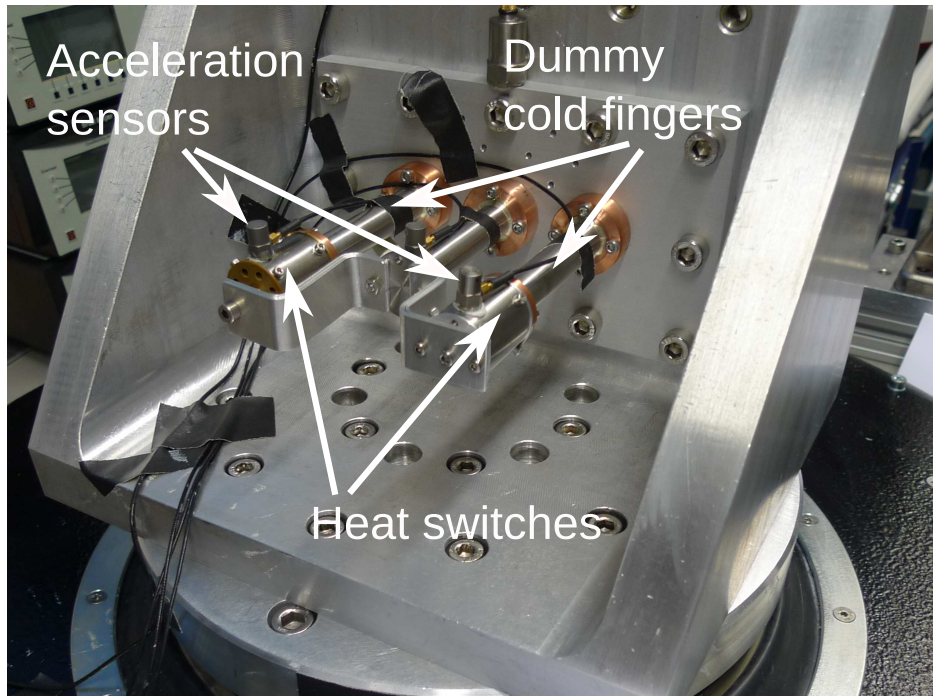}
\par\end{centering}
\caption{Test rig with two heat switches on the shaker (photo courtesy of AIM
GmbH).\label{fig:Photo-of-the-1}}
\end{figure}
 (see Fig. \ref{fig:Photo-of-the-1}). The difference of the response
curve at the beginning and at the end of the test showed almost no
change, which proves that no structural damage occurred during this
test. In addition, the switch thermal performance measured before
and after the shaker test did not reveal any degradation.

The long-term stability was tested by measuring the temperature difference
over the switch in closed state over a period of 6 months
\begin{figure}
\begin{centering}
\includegraphics[width=0.8\textwidth]{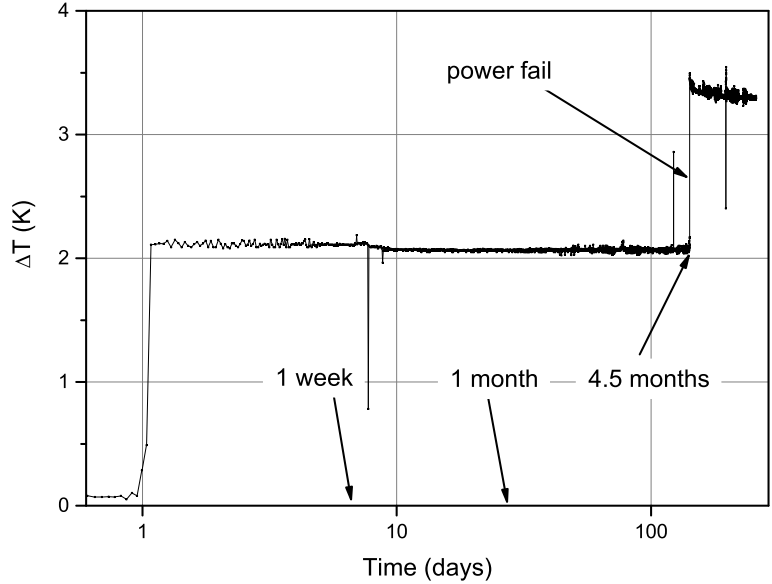}
\par\end{centering}
\caption{Long term measurement of the temperature difference across the switch
of Ref. \cite{Dietrich2014} at a heat load of 1 W. Cold flange temperature
$T_{CF}$= 100 K.\label{fig:Long-term-measurements.}}
\end{figure}
 (see Fig. \ref{fig:Long-term-measurements.}). For this test, the
switch from the previous project \cite{Dietrich2014} was used, which
already underwent several dozens of cooling/warming cycles. Therefore,
the switch can be regarded as ``pre-aged''. In Fig. \ref{fig:Long-term-measurements.},
the temperature difference $\Delta T$ across the switch was monitored
at a constant heat load of 1 W applied to the sensor side of the switch
shaft. As seen from the Figure, $\Delta T$ is stable or even decreases
a little bit for a period of more than 4 months. After 4.5 months,
a power outage caused a vacuum pump failure and the switch warmed
up at the same time. After power was restored, the switch was not
able to return to its previous on-state performance. This behaviour
may be caused by a contamination of the contact surface due to the
vacuum leak. Besides this change in on-state performances, the switch
showed no signs of degeneration over the initial 4 month of testing.
A subsequent short term measurement over one month, which included
a sudden heat-up to 200 K, showed no degeneration of the on-state
conduction after the heat switch was cooled down to 100 K again. This
means that pre-aging of the actuator material is an option to maintain
stable long-term operation of this kind of thermal switch.

Another method to avoid creeping is to thermally anneal the thermoplastic
material after machining has been finished. To evaluate this option,
the UHMW-PE actuator of the newly designed switch was replaced by
one, which was annealed at different temperatures. The switch was
first cooled under its closing temperature and warmed up to room temperature
again. This process was repeated several times and the closing/opening
temperatures were monitored. When the actuator material was annealed
using the UHMW-PE supplier's recommended thermal cycle between room
temperature and 85\textdegree C, it showed severe degeneration in
closing temperature as can be seen from 
\begin{figure}
\subfloat[Annealing temperature 85\textdegree C.]{\begin{centering}
\includegraphics[width=0.5\textwidth]{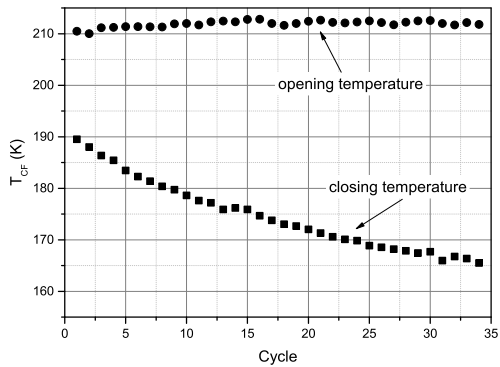}
\par\end{centering}
}\subfloat[Annealing temperature 105\textdegree C.]{\begin{centering}
\includegraphics[width=0.5\textwidth]{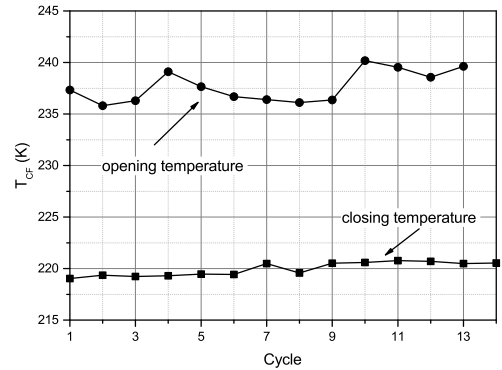}
\par\end{centering}
}

\caption{Open/Close-cycles of the new switch with different annealing temperatures
of the UHMW-PE actuator.\label{fig:Open/Closing-cycles-with-an}}
\end{figure}
Fig. \ref{fig:Open/Closing-cycles-with-an}a. The annealing temperature
was then increased from 85\textdegree C to 105\textdegree C using
a new actuator sample and this time the switch showed almost no change
in opening/closing temperatures as shown in Fig. \ref{fig:Open/Closing-cycles-with-an}b.
As expected, the on-state condition after the cyclic actuation stayed
at its initial value of 250 mW/K at the beginning of the test cycle.
Proper annealing is therefore an important process for gaining repeatable
switch operation.

\section{Conclusions}

Using a combined thermo-mechanical model, the mass and the stability
of a CTE-based heat switch was optimized. The mass of the new switch
is only 25 g compared to 250 g of the previously realized switch \cite{Dietrich2014}.
The calculated thermal conductances are compared with experimental
data. The predicted off-state conductance is higher than in the experiment
due to additional contact resistances which are not included in the
model. The predicted on-state conductance is lower than in the model,
likely because of the imperfect contact area alignment. This gives
rise for further improvement. The developed models enable the easy
scaling of the switch dimensions to fulfil the requirements by a given
application.

In contrast to the previous development, the mechanical and thermal
stability of the heat switch is proven by shaker tests and long-term
measurements. The next step will be the incorporation of the switch
design in a practical space application.

\section*{Acknowledgements}

This work was financially supported by the German Federal Ministry
of Economics and Technology (BMWi Grant No. 50 EE 1322). The authors
thank AIM Infrared Modules GmbH (Heilbronn) for performing the shaker
tests. Useful discussions with Jan Grosser and Alexander Gerling (DLR
Bonn) are gratefully acknowledged. 

\section*{References}

\bibliographystyle{elsarticle-num}
\addcontentsline{toc}{section}{\refname}\bibliography{thermoswitch}

\end{document}